# Evaluating the effectiveness of concrete and metal shields in photoneutron reduction: A Monte Carlo study

M H Jahanimehr[a,*], S M M Abtahi[a]

[a]Departement of physics, Imam Khomeini International University, Qazvin, Iran

**Abstract**

**Introduction:** Neutrons are produced when photons interact with parts of the accelerator, materials in the treatment room, or the patient's body. These neutrons have significant biological effects due to their high quality factor, potentially leading to secondary cancers.

**Material and method:** A Monte Carlo simulation was conducted using the MCNPX code to model an Elekta Infinity linear accelerator operating at 18 MV. The accelerator was situated inside a bunker measuring 7.4×5 m$^2$. To study photoneutron production and the effectiveness of shielding, various concrete mixtures were applied to the walls of the maze and treatment room.

**Results:** Regardless of the modifications, both photon and neutron fluence decreased substantially as the distance from the maze entrance increased. The fluence reduction was more pronounced between the maze entrance and center. The lowest equivalent neutron doses behind the treatment room door were observed when Ord-Lim concrete was used, while the highest doses occurred with Gal-Bar concrete.

**Conclusion:** The configuration of concrete layers within the treatment room and maze walls is a key factor influencing the changes in neutron dose behind the treatment room door. These variations result from differences in concrete composition and the way neutrons are absorbed in different setups.

**Keywords:** Elekta Infinity 18 MV, photon fluence , neutron fluence , concrete shields

**Highlights**

1. Ord-Lim produces a lower neutron dose behind the room compared to other configurations.
2. The arrangement and order of different types of concrete in the bunker walls have a significant impact on the production rate of photoneutrons.
3. The use of metals inside and outside the room has a greatly influences on neutron production.
4. Placing a layer of steel between two layers of concrete plays an effective role in reducing photoneutron production.

## 1. Introduction

Neutrons are generated when high-energy photons and electrons interact with dense materials found in accelerator components, such as the target and collimators [1]. Gamma radiation can also interact with the treatment room components and the patient's body, resulting in the generation of photoneutrons [2]. Because of their high radiobiological effectiveness, neutrons can cause secondary cancers in patients. The total neutron fluence ($\varphi_T$) is the sum of direct neutrons ($\varphi_d$), scattered neutrons ($\varphi_{sc}$), and thermal neutrons ($\varphi_{th}$), as shown in Equation 1.1[3].

$$\varphi_T = \varphi_d + \varphi_{sc} + \varphi_{th} \tag{1.1}$$

Multiple studies have measured the neutron equivalent dose at the isocenter, different locations within the maze, behind walls, and at the entrance door using both simulation and experimental approaches. The primary purpose of determining the neutron equivalent dose is to assess the biological impact of neutrons in radiobiology[4]. In numerous other studies, the neutron equivalent dose has been determined using a technique referred to as the Kersey method [5, 6].

$$H = H_0 \frac{T}{T_0} \frac{d_0^2}{d_1^2} 10^{-d_2^{2/5}} \tag{1.2}$$

In Equation 1.2, $H_0$ represents the neutron equivalent dose measured at the isocenter; $d_0$ is the distance from the target to the isocenter; $d_1$ is the distance from the isocenter to the inner maze entrance; $d_2$ is the distance from the maze center to the wall; and $T/T_0$ denotes the ratio of the smallest to the largest cross-sectional areas of the maze.

Researchs have been carried out and strategies suggested to minimize photoneutron production. Using the Geant4 simulation code, Moghaddasi and Colyer showed that layered configurations of concrete and steel can effectively reduce neutron levels[7]. They also reported that using steel as both the primary and secondary shielding layers, with average thicknesses of 16.6 cm and 33.3 cm respectively, for an Elekta 18 MV accelerator, can effectively reduce photoneutron production. However, another study using the MCNPX code for a Varian 2100/2300 C/D 18 MV accelerator found that employing lead and steel shields was not effective in decreasing photoneutron levels[8]. Lead has a low tenth-value layer (TVL) for photons, but its neutron emission threshold energy is relatively low (6.74 MeV). As a result, it is not an appropriate material for neutron shielding [9]. Steel has a neutron emission threshold energy of 11.2 MeV, indicating that it generates fewer photoneutrons than lead [9]. However, some studies have noted that since treatment room space is often limited, using these two types of barriers can help decrease the required thickness of shielding walls [10, 11]. In a separate study, Mesbahi et al. investigated various types of concrete and found that heavy concretes can reduce the amount of primary and secondary shielding needed in bunkers. This is due to the fact that these concretes contain dense minerals such as limonite, barite, and magnetite, along with high atomic number materials like lead and iron[12]. They tested different types of light and heavy concrete for the bunker walls and calculated the neutron dose at the maze entrance for each concrete type. In their study, the materials used for the treatment room walls and the maze were the same. Similarly, Ghiyasi and Mesbahi explored various concrete types to reduce photoneutrons. They simulated a Varian 2100 accelerator and its bunker, calculated neutron fluence at both the maze entrance and exit, and found that heavy concretes such as magnetite-steel and limonite-steel generate more photoneutrons in the treatment room compared to lighter concretes like ordinary and serpentine[13]. In this study, we aim to introduce a new method utilizing heterogeneous concretes. Specifically, the treatment room walls were constructed with heavy concrete, while the maze walls were built with light concrete, and the reverse arrangement was also examined.

## 2. Material and method

An Elekta 18 MV linear accelerator was modeled using the Monte Carlo MCNPX code. The accelerator included components such as the target, primary collimator, secondary collimator, ion chamber, multi-leaf collimator (MLC), flattening filter, and scattering foil. These parts were composed of metals like tungsten, iron, and tantalum. A visual representation of the accelerator is provided in Figure 1.

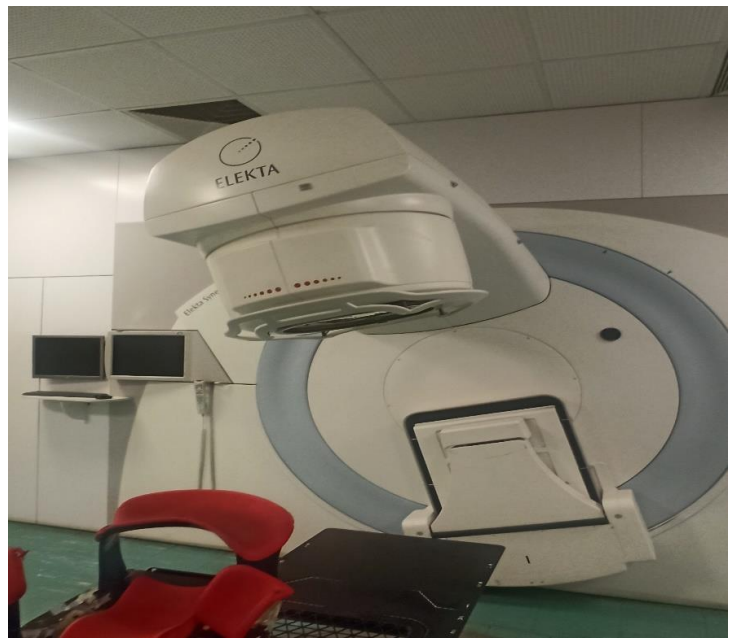

Figure1. A perspective of the Elekta 18 MV radiation accelerator

### 2.1. Validation

A water phantom measuring 40 × 40 × 40 cm³ was positioned 100 cm away from the target. The field size was set to 5 × 5 cm². The phantom was segmented into cylindrical meshes with a radius of 1.25 cm and a height of 1.25 mm, and the photon energy deposition in each mesh was recorded using the Type 3 tally mesh card. The percentage depth dose (PDD) at 10 cm depth in the water phantom was normalized, and the results were compared with experimental data from the study by Almberg et al.[14]. The greatest difference between the two curves was approximately 5% at the graph's peak, while the smallest difference was around 2%. This indicates that our results closely matched the experimental data.

### 2.2. Bunker geometry

In this study, a bunker comprising a treatment room and a 10-meter-long maze was simulated, with the bunker geometry modeled similarly to that described by Mesbahi et al[12]. The treatment room measured approximately 7.4 m × 5 m, with a ceiling height of 10 meters. The maze area was 10 meters long and 1.75 meters wide, and the concrete wall thickness ranged from 1.5 to 2.5 meters. Different types of concrete were used for the treatment room walls and the maze walls. Concretes with a density below 4 g/cm³ were classified as light or medium, while those with a density above 4 g/cm³ were considered heavy concretes. Figure 2a provides an illustration of the accelerator bunker layout.

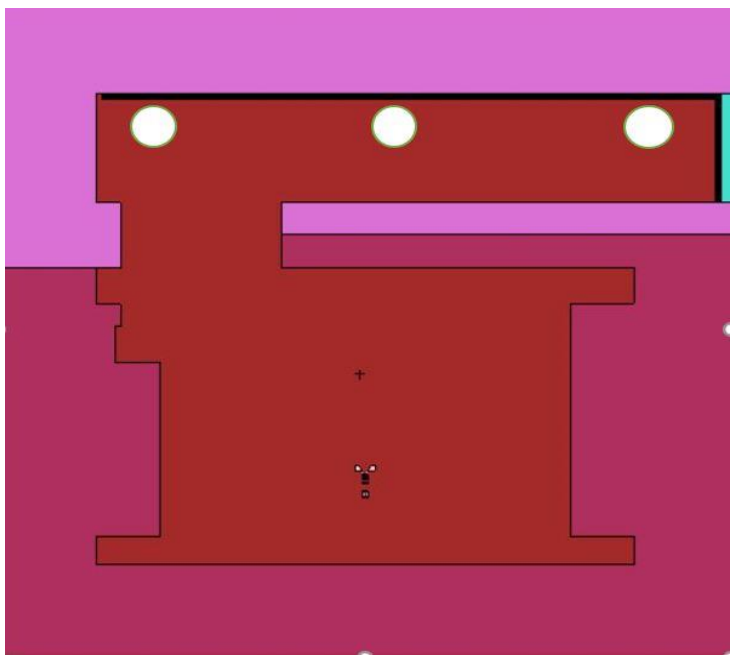

Figure 2a. Use of different concretes in the maze and treatment room walls

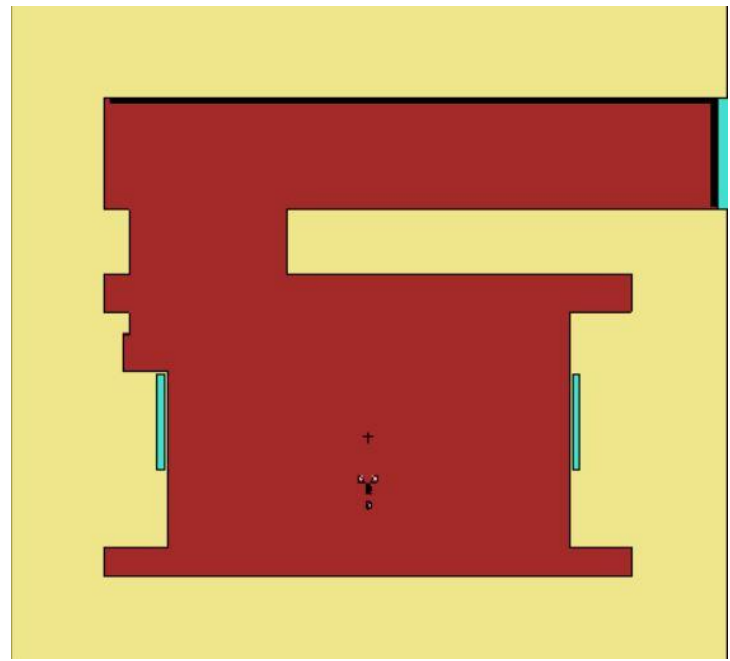

Figure 2b. Use of a 1 TVL steel barrier in the walls parallel to the accelerator

The selection of concretes placed side by side was based on maximizing the difference in density, with concretes having the largest density contrast (such as Serpentine and Magnetite-steel concretes) paired together, while those with smaller density differences were positioned adjacent to each other. Six configurations were considered:

- Bar-Gal concretes: Treatment room walls made of barite concrete, maze walls made of galena concrete.
- Gal-Bar concretes: Treatment room walls made of galena concrete, maze walls made of barite concrete.
- Mag-Ser concretes: Treatment room walls made of magnetite-steel concrete, maze walls made of serpentine concrete.
- Ser-Mag concretes: Treatment room walls made of serpentine concrete, maze walls made of magnetite-steel concrete.
- Ord-Lim concretes: Treatment room walls made of ordinary concrete, maze walls made of limonite-steel concrete.
- Lim-Ord concretes: Treatment room walls made of limonite-steel concrete, maze walls made of ordinary concrete.

Table1 shows the density and compositions of the concretes used [15-17].

Table1. Researched concrete materials and their constituent components extracted by NCRP 144 and the studies of Mortazavi et al.

| Concrete types / Elements | Barytes 3.35 $\frac{g}{cm^3}$ | Datolite galena 4.42 $\frac{g}{cm^3}$ | Magnetite-steel 4.64 $\frac{g}{cm^3}$ | Serpentine 2.1 $\frac{g}{cm^3}$ | Ordinary 2.35 $\frac{g}{cm^3}$ | Limonite-steel 4.54 $\frac{g}{cm^3}$ |
|---|---|---|---|---|---|---|
| Hydrogen | 0.012 | 0.039 | 0.011 | 0.035 | 0.013 | 0.031 |
| Oxygen | 1.043 | 0.770 | 0.638 | 1.126 | 1.165 | 0.708 |
| Silicon | 0.035 | 0.310 | 0.073 | 0.460 | 0.737 | 0.067 |
| Calcium | 0.168 | 0.509 | 0.258 | 0.15 | 0.194 | 0.261 |
| Carbon | --- | --- | --- | 0.002 | --- | --- |
| Sodium | --- | 0.008 | --- | 0.009 | 0.04 | --- |
| Magnesium | 0.004 | --- | 0.017 | 0.297 | 0.006 | 0.007 |
| Aluminum | 0.014 | 0.013 | 0.048 | 0.042 | 0.107 | 0.029 |
| Sulfur | 0.361 | 0.357 | --- | ---- | 0.003 | --- |
| Potassium | 0.159 | 0.010 | --- | 0.009 | 0.045 | 0.004 |
| Iron | --- | 0.028 | 3.512 | ---- | 0.029 | 3.421 |
| Titanium | --- | --- | 0.074 | 0.002 | --- | --- |
| Chromium | --- | 0.005 | --- | --- | --- | --- |
| Manganese | --- | 0.006 | --- | --- | --- | --- |
| Vanadium | --- | --- | 0.003 | --- | --- | 0.004 |
| Barium | 1.551 | --- | --- | --- | --- | --- |
| Lead | --- | 2.303 | --- | --- | --- | --- |
| Boron | --- | 0.046 | --- | --- | --- | --- |

Another bunker, with the same dimensions shown in Figure 2a, was also simulated, but with walls made of standard concrete (Figure 2b). One tenth-value layer (TVL) of steel and lead was incorporated into the walls parallel to the accelerator, and the equivalent neutron dose behind the treatment door was measured. Additionally, to assess the impact of moderators on neutrons, a 4 cm thick polyethylene board (BPE) was positioned behind the treatment door, and the equivalent neutron dose was calculated.

## 2.3. Simulation using the Monte Carlo technique

The Monte Carlo code MCNPX 2.6.0 was utilized to simulate photon and neutron transport. Neutron tracking was enabled using the physics card. The electron beam energy for the Elekta Infinity device was set to 18 MeV as a pencil beam. Photon and neutron fluences were calculated at one meter above the room floor, specifically at the corner, center, and end of the maze, using the F4 tally. The equivalent neutron dose behind the treatment door was determined with the DF. In this simulation, $1\times10^7$ electrons were directed at the target. To minimize statistical error, the Russian roulette method was applied near the room door. The Cut card was used to decrease computation time and enhance simulation efficiency. Since most accelerator components have a neutron emission threshold above 7 MeV, a cutoff energy of 7 MeV was chosen for photons, meaning particles were terminated when reaching this energy and the simulation continued with the next particle. The program was executed on a computer with 16 GB RAM, a 64-core processor, and a 4th generation CPU, with a total runtime of about two months.

## 3. Results

Figure 3a displays a comparison of the measured and simulated depth dose distributions in water along the beam's central axis. The results indicate a close agreement between the measured and calculated depth dose profiles.
Figure 3b illustrates the differences between the two approaches.

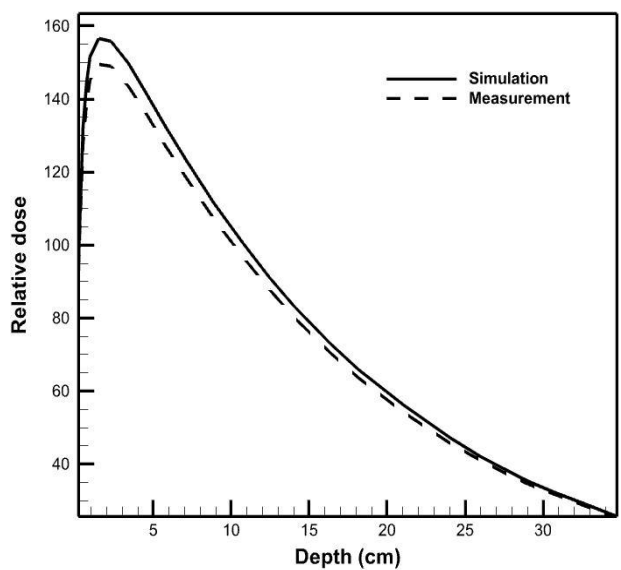

Figure 3a. Comparison of the percentage depth dose curve with experimental results

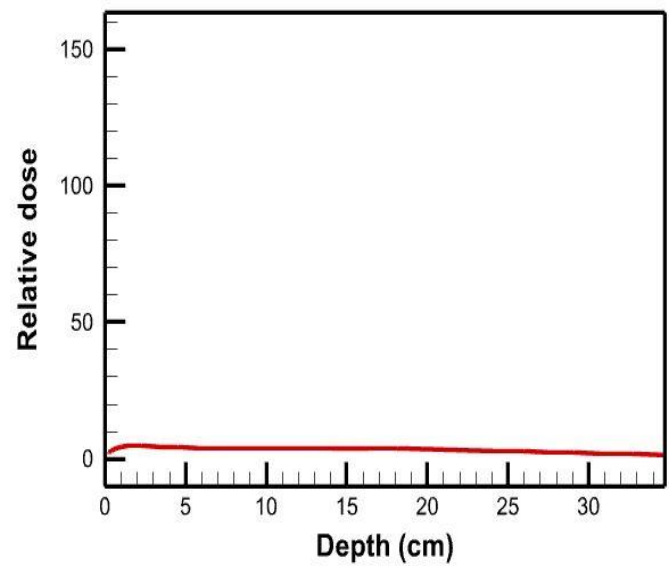

Figure 3b. Difference between experimental and simulation results

In this study, neutron and photon fluences were calculated at the maze entrance, center, and exit. The simulation results showed that the highest neutron fluence near the maze entrance was observed in the Lim-Ord configuration, while the lowest was in the Bar-Gal configuration. Near the maze exit, the highest neutron fluence was found in the Gal-Bar configuration, and the lowest in the Bar-Gal configuration. The photon fluence along the maze was also calculated at different points. The highest photon fluence near the maze entrance was associated with the Mag-Ser configuration, while the lowest was associated with the Gal-Bar configuration. Figures 4a and 4b illustrate the neutron and photon fluences, respectively, at various distances along the maze.

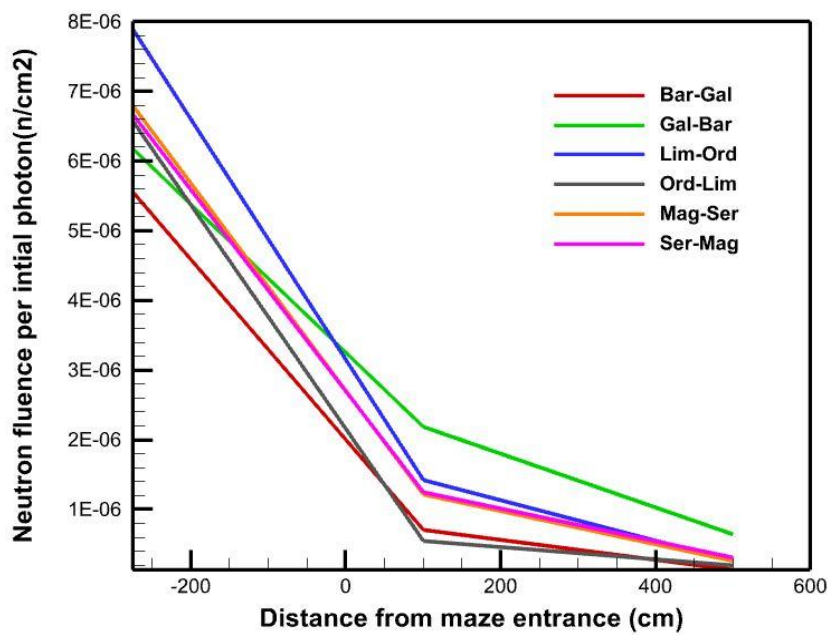 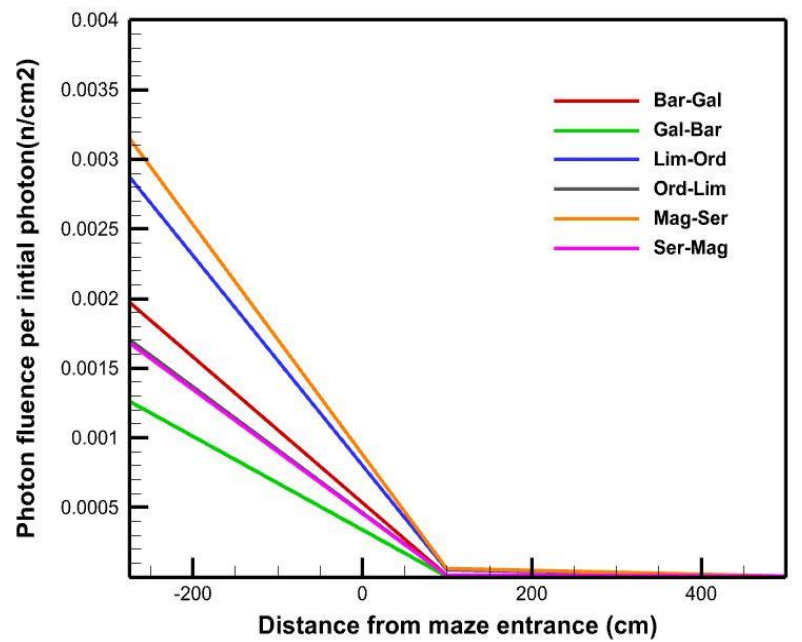

Figure 4a: neutron fluence at different distances from the maze

Figure 4b: photon fluence at different distances from the maze

As illustrated in Figure 4b, the photon fluence near the maze exit is roughly similar across all cases. The photon fluence curve shows a steep decline from the maze entrance to the maze center, followed by only minor changes thereafter. Similarly, neutron fluence drops sharply from the maze entrance to the center and then decreases more gradually from the center toward the entrance door. Maglieria et al., based on experimental data, reported that neutron fluence at the maze center is reduced by 94% compared to the maze entrance[18].

The equivalent neutron dose behind the treatment door was determined for all six configurations. The results indicated that the Ord-Lim case had the lowest equivalent neutron dose, while the Gal-Bar case had the highest. Table 2 summarizes the equivalent neutron dose values for each of the six cases.

Table2. Neutron dose equivalent behind the door (mSv/Gy)

| | |
|---|---|
| Bar-Gal | $1.24 \times 10^{-3}$ |
| Gal-Bar | $1.93 \times 10^{-3}$ |
| Lim-Ord | $1.04 \times 10^{-3}$ |
| Ord-Lim | $8.84 \times 10^{-4}$ |
| Mag-Ser | $1.01 \times 10^{-3}$ |
| Ser-Mag | $1.27 \times 10^{-3}$ |

Our findings also revealed that incorporating steel and lead increased the neutron equivalent dose behind the door by 30% and 61%, respectively. Furthermore, using BPE reduced the neutron equivalent dose by a factor of 5.3.

## 4. Discussion

In study Mesbahi et al., the neutron fluence near the maze entrance per primary photon was reported as $1.90 \times 10^{-5} \frac{n}{cm^2}$ on average for various concrete types[12]. In contrast, our results showed an average neutron fluence of $6.60 \times 10^{-6} \frac{n}{cm^2}$ at the same location. The reason for the difference in results between the two studies could be that in our study, two types of concrete were used in designing the bunker walls, whereas in the study by Mesbahi et al., only one type of concrete was simulated for the bunker walls. Another reason for the discrepancy between the two studies could be that they used a Varian accelerator, while in this study was used an Elekta Infinity accelerator. In the study by Carinou et al., barite concrete was used for the bunker walls. They reported a neutron dose at the maze entrance that was considerably higher than what this study observed. This difference can be attributed to the maze dimensions: in our study, the maze length and width were 10 meters and 1.75 meters, respectively, while in their study, these measurements were 7.5 meters and 2.5 meters, respectively [19]. They demonstrated using the MCNP code that increasing the maze length from 6.5 meters to 9.5 meters can reduce the neutron dose at the maze entrance by up to 80%, while decreasing the maze width from 2 meters to 1.5 meters results in approximately a 25% reduction in neutron dose. Therefore, the notable difference between our study and theirs appears justified. Moghaddasi and Colyer employed the Geant4 code along with Varian accelerators of 10 and 18 MV and an Elekta 18 MV accelerator. They positioned a water phantom 30 cm away from the concrete barrier and compared the neutron transmission ratio to a scenario where a steel barrier was used in combination with concrete[7]. Their findings revealed significantly lower neutron transmission at a 1 cm depth in the water phantom when steel barriers were combined with concrete, compared to pure concrete. For Elekta and Varian 18 MV accelerators, incorporating 8 cm of steel alongside concrete reduced neutron transmission by approximately 70% relative to concrete alone. They highlighted that despite iron's low hydrogen content, its high elastic and inelastic neutron scattering cross-section in the MeV energy range enhances shielding efficacy.

In another study, Mendes et al. simulated an 18 MV Varian accelerator inside a bunker with standard concrete walls. Their results showed that substituting lead with steel in the walls parallel to the accelerator lowered the neutron dose at the patient's treatment couch by about 10%. However, they did not directly compare these findings to cases where only concrete walls were used[8]. Choi et al. used the MCNPX code to study how different barriers affect neutron production rates. They positioned a 1.5-meter-thick concrete barrier 4.5 meters from the source, behind a steel barrier with one TVL thickness, and calculated the equivalent neutron dose at 50 cm from the barrier. Their findings showed that the presence of the steel barrier increased the neutron dose by a factor of 20 [20]. In their study, they replaced the steel barrier with lead and found that the presence of lead increased the neutron dose by a factor of 50 compared to using only concrete. They also experimented with placing 22 cm of concrete on both sides of the metal barrier, effectively sandwiching the metal between two concrete layers, which resulted in a significant reduction in the equivalent neutron dose. Facure et al., using a 10 MV accelerator, demonstrated that a lead barrier causes a substantial increase in neutron dose both inside and outside the treatment room. Their research also showed that positioning lead sheets behind the concrete generates more neutrons than placing them in front. Moreover, they reported that using a steel barrier instead of concrete does not present a neutron hazard[11]. McGinley investigated the effects of varying thicknesses of lead and steel both inside and outside bunker walls for 18 MV Varian and Siemens accelerators. The study found that substituting steel for lead reduced photoneutron production by a factor of 11. However, the presence of either metal as shielding on both the interior and exterior of the bunker led to a substantial increase in neutron dose [21]. Martínez-Ovalle et al. assessed neutron dose from an 18 MV Varian accelerator installed in a complex bunker with walls constructed of both ordinary and barite concrete. They measured a neutron dose of 12.5 mSv per hour near the upper door gaps. Their simulations of various bunker geometries and accelerator types further showed that neutron fluence outside the room door is unavoidable [22]. In the study by Zabihzadeh et al., a bunker comprising a treatment room and two mazes was simulated. Their findings indicated that the equivalent neutron dose increases as the distance from the inner maze entrance door grows. Specifically, the neutron dose near the maze entrance was found to be five times greater than that near the inner maze door. They also noted that fast neutrons are absent inside the maze because the maze walls slow them down through interactions, effectively thermalizing them[23]. In the study by Falcao et al., three different bunker designs housing 18 MV accelerators were simulated using the MCNP code. Their results demonstrated that neutron fluence at various locations within the maze is highly influenced by the bunker's geometry. In two of the bunkers, the neutron fluence at the maze center was twice as high as at the maze end, while in the third bunker, it was 70% higher at the maze center compared to the maze end. Furthermore, in all three cases, the difference in neutron fluence between the maze entrance and the maze center was greater than the difference between the maze center and the maze end, aligning with the findings of our study[5]. Table 3 presents a comparison between the findings of this study and those of other research. Our study demonstrated that Bar-Gal generates lower neutron fluence throughout the maze compared to other types of concrete. While barite alone is not highly effective at attenuating thermal and fast neutrons, combining it with galena can improve the overall neutron attenuation performance[24]. Another factor contributing to the neutron attenuation observed with Bar-Gal concrete is barite's effectiveness in attenuating gamma rays. When used in the treatment room walls, barite substantially lowers photon fluence, which in turn reduces the probability of photoneutron production [25]. Our study revealed that using Gal-Bar (the reverse arrangement of Bar-Gal) results in the highest neutron fluence near the maze entrance compared to other configurations. This highlights that the sequence in which different concretes are placed next to each other significantly influences the rate of photoneutron production. Furthermore, our findings showed that neutron fluence near the maze entrance is greater for the Lim-Ord and Mag-Ser setups than for the Ord-Lim and Ser-Mag configurations. This suggests that employing heavy concretes in the treatment room walls combined with lighter concretes in the maze walls leads to increased photoneutron production. Heavy concretes, composed of dense minerals with high atomic numbers and low hydrogen content, are less effective at elastically scattering neutrons, which results in higher neutron fluence at the maze entrance. Additionally, our study indicated that concretes with a high iron content-such as those containing magnetite and steel or limonite and steel-when used for treatment room walls, generate the highest photon fluence at the maze entrance. Photon fluence at the maze entrance was higher in the Gal-Bar configuration compared to Bar-Gal. This is likely because datolite galena contains a high concentration of lead, and due to lead's high atomic number and electron density, it has a stronger ability to scatter and absorb gamma rays than barite. The study's results also indicated that the reduction rate of photon and neutron fluence with distance is greater for the Mag-Ser and Lim-Ord configurations than for the other setups, possibly because serpentine and ordinary concretes have lower densities. Since light concretes contain less lead and iron than heavy concretes, their use is more effective in decreasing the photoneutron production rate[13]. Additionally, to further decrease the equivalent neutron dose, borated polyethylene was applied near the entrance door, resulting in a reduction of the neutron dose by a factor of 5.3. Similarly, Krmar et al. used a paraffin sheet placed one meter from the maze entrance, which led to a threefold decrease in the equivalent neutron dose[26]. In this study, borated polyethylene achieved a greater reduction in neutrons because it contains boron, which has a neutron absorption cross-section exceeding 3800 barns. In contrast, materials like paraffin and polyethylene primarily serve as neutron moderators and have relatively low neutron absorption cross-sections[27]. Carinou and et al. applied a 1-cm-thick layer of borated polyethylene to cover the inner surfaces of the walls, reporting a 47% reduction in neutron dose at the bunker door entrance[19]. Comparing their findings with those of our study suggests that placing moderating materials like polyethylene and paraffin closer to the entrance door leads to more effective neutron reduction. This is likely because the average neutron energy decreases rapidly along the maze, making neutron attenuation by moderating materials near the door significantly more efficient.

Table3. Comparison of neutron dose in the presence of lead, steel, and BPE with other studies

| Source | Method | Neutron dose | Measurement point | References |
|---|---|---|---|---|
| Elekta Infinity 18MV | Monte Carlo code MCNPX | $1.03\times10^{-3}$ mSv/Gy (ordinary concrete)<br>$1.34\times10^{-3}$ mSv/Gy (With the steel barrier)<br>$1.66\times10^{-3}$ mSv/Gy (With the lead barrier)<br>$0.19\times10^{-3}$ mSv/Gy (With BPE) | Outside the room<br>Outside the room<br>Outside the room<br>Outside the room | This study |
| High energy photon | Monte Carlo code Geant4 | $5.2\times10^{-8}$ Gy (with the steel barrier)<br>$2.1\times10^{-6}$ Gy (Pure concrete) | 1cm depth in water<br>1cm depth in water | [7] |
| Varian 2100/2300 C/D 18 MV | Monte Carlo code MCNPX | 2.2 mSv/Gy (with the steel barrier)<br>2.5 mSv/Gy (with the lead barrier) | Patient's plane<br>Patient's plane | [8] |
| Varian CL21EX 18 MV | nested neutron spectrometer | 54.1 mSv/h (Ordinary concrete)<br>1 mSv/h (Ordinary concrete) | Maze entrance<br>Middle of the maze | [18] |
| Saturn 20 CGR 18Mev | Monte Carlo code MCNP4C | $3.3\times10^{-2}$ mSv/Gy (Low density concrete)<br>$5.5\times10^{-3}$ mSv/Gy (Low density concrete) | 20 cm from end of maze<br>380cm from end of maze | [23] |
| Varian 21iX linac 18MV | Monte Carlo code MCNPX and EGSnrc/BEAMnrc | 5 mSv/Gy (Lead barrier infront concrete)<br>0.2 mSv/Gy (Lead barrier sandwiched between two layers of concrete)<br>0.09 mSv/Gy (Steel barrier sandwiched between two layers of concrete) | 50 cm from the inner wall<br>50 cm from the inner wall<br>50 cm from the inner wall | [20] |
| Varian 18 MV | Monte Carlo code MCNP | $6.35\times10^{-4}$ mSv/Gy (Ordinary concrete)<br>$2.50\times10^{-3}$ mSv/Gy (Ordinary concrete)<br>$2.57\times10^{-2}$ mSv/Gy (Ordinary concrete) | At the door position<br>At the door position(Changing bunker geometry)<br>At the door position(Changing bunker geometry) | [5] |
| 10MV Medical accelerator | Monte Carlo code MCNP | 1.56 μSv (1 TVL Lead infront ordinary concrete)<br>1 μSv (2 TVL Lead infront ordinary concrete)<br>0.91 (4 Tvl Lead infront ordinary concrete) | Outside the room<br>Outside the room<br>Outside the room | [11] |
| 18 MV Saturne Medical linac | TLD measurement | 0.21 mSv/Gy (ordinary concrete)<br>0.018 mSv/Gy (ordinary concrete) | Maze entrance<br>Corner of the maze | [28] |
| Varian 2100C 15MV | A Meridian model 5085 neutron survey meter | $0.57\times10^{-3}$ mSv/Gy | near the maze entrance door | [26] |
| Varian Clinac 23EX 18MV | NP-2 Rem meter | 925 μSv h$^{-1}$ (before BPE)<br>550 (after BPE) | Maze entrance<br>Maze entrance | [29] |
| Philips SL 18MV | Monte Carlo code MCNP | $3.80\times10^{-4}$ mSv/Gy (NBS-5 concrete)<br>$1.30\times10^{-4}$ mSv/Gy (NBS-5 + 1cm BPE)<br>$3.12\times10^{-4}$ mSv/Gy (Barite concrete) | Door entrance<br>Door entrance<br>Door entrance | [19] |
| Varian2100 C/D 18MV | Monte Carlo code MCNPX | $5.63\times10^{-2}$ mSv/Gy (Hematite concrete)<br>$6.5\times10^{-2}$ mSv/Gy (Hematite concrete)<br>$9.48\times10^{-4}$ mSv/Gy (Hematite concrete) | Maze entrance<br>Middle entrance<br>Entrance door | [30] |
| Varian Clinac 2100 18MV | Monte Carlo code MCNPX | $1.57\times10^{-3}$ mSv/Gy (ordinary concrete)<br>$2.1\times10^{-3}$ mSv/Gy (Magnetite concrete)<br>$2\times10^{-3}$ mSv/Gy (Barite concrete)<br>$2.5\times10^{-3}$ mSv/Gy (Magnetite steel concrete)<br>$2.46\times10^{-3}$ mSv/Gy (Limonite steel concrete)<br>$1.5\times10^{-3}$ mSv/Gy (Serpentine concrete) | Bunker entrance<br>Bunker entrance<br>Bunker entrance<br>Bunker entrance<br>Bunker entrance<br>Bunker entrance | [13] |

## 5. Conclusion

The study conducted showed that the Ord-Lim configuration produces a lower equivalent neutron dose behind the treatment room door compared to other configurations, indicating that the composition and structure of different concretes affect protective and nuclear properties. Additionally, the effect of lead and steel barriers on neutron dose was examined and compared with other studies. The results demonstrated that optimizing the combination of metal and concrete in protective structures not only improves radiation safety but also reduces construction costs. Using a sandwich arrangement of steel between two concrete layers reduces the neutron dose, whereas placing barriers in front of or behind the concrete causes a significant increase in photoneutron production. It cannot yet be conclusively determined how sandwiching lead between two layers of concrete affects the neutron production rate. Using 1 TVL of lead and steel in front of concrete walls increased the equivalent neutron dose by 61% and 30%, respectively. The effect of borated polyethylene on reducing neutron dose was investigated, and it was shown that using borated polyethylene loaded onto the treatment room door reduces the neutron dose by a factor of 5.3. However, increasing the distance of the polyethylene sheet from the door decreases its effectiveness. Additionally, the discrepancies in neutron dose calculations in different studies were attributed to differences in bunker dimensions, especially the maze, as well as the type of accelerator and the type of concrete used in the treatment room walls.

**Acknowledge**

The Oncology Unit at Mahdieh Diagnostic and Treatment Center, Hamadan is sincerely appreciated by the authors for their support of this project.